\documentclass[12pt,a4paper]{aastex631}
\shorttitle{Clumpy accretion in p-m-s stars as a source of perturbations in circumstellar disks}
\shortauthors{Demidova \& Grinin}
\graphicspath{{./}{figures/}}

\begin{document}

\title{Clumpy Accretion in Pre-main-sequence Stars as a Source of Perturbations in Circumstellar Disks}

\correspondingauthor{Tatiana V. Demidova}
\email{proxima1@list.ru}

\author[0000-0001-7035-7062]{Tatiana V. Demidova}
\affiliation{Crimean Astrophysical Observatory,p. Nauchny, Bakhchisaray, Crimea, 298409}

\author[0000-0001-8923-9541]{Vladimir P. Grinin}
\affiliation{Pulkovo Observatory of the Russian Academy of Sciences, Pulkovskoje Avenue 65, St. Petersburg 196140, Russia}
\affiliation{V.V. Sobolev Astronomical Institute, St. Petersburg University, Petrodvorets, St. Petersburg, Russia}



\begin{abstract}
The development of perturbations in the circumstellar disks of pre-main-sequence stars caused by clumpy accretion was investigated. Here we perform 3D hydrodynamical SPH simulations of disks perturbed by a recent clump accretion event. These simulations are further explored by radiative transfer calculations to quantify observational appearance of such disks. It was shown that the density waves in the disks were formed at the fall of the clump.  After  several revolutions they can transform  into spirals and ring structures.  Their images in mm-wavelengths are very similar to those observed with ALMA in some protoplanetary disks. We assume that the clumpy accretion may be the source of such structures.

\end{abstract}



\section{Introduction} \label{sec:intro}

After formation of a young star and its circumstellar disk due to the gravitation collapse of the protostellar cloud the process of accretion of matter from the nearest surrounding the star can continue and have a form of the clumpy accretion. \citet{1992PASP..104..479G} was probably the first who used this term. He tried to explain by such a way  the observations of the strong extinction events observed in  some young variables. Later this type of accretion has been considered by \citet{1996ARA&A..34..207H} for explanation of the FUOR’s  phenomenon.  This idea is quite popular up to now \citep{2010ApJ...713.1143Z, 2013ApJ...764..141B, 2018MNRAS.474...88H}. The mechanism of formation of chondrule as a result of the clumpy accretion was discussed  \citet{1998Icar..134..137T}. Obviously, the fall of the clump on the disk should cause disturbances at the place of the fall. It is interesting to trace how this disturbance will develop and what structures on the disk can be caused by it.
 
The detection of various structures on images of protoplanetary disks is one of the most interesting results obtained with the ALMA interferometer \citep[see, e.g.][]{2018A&A...619A.161C, 2018ApJ...869L..43H,2018ApJ...869L..42H, 2018ApJ...869L..50P}. Ring and spiral structures are most often observed. Less commonly, structures resembling highly elongated vortices are observed. A number of papers are devoted to theoretical studies of the formation of such structures. Their formation is associated with perturbations in the disks caused by the motion of the forming planets \citep[e.g.,][]{2013A&A...549A..97R, 2015A&A...579A.106V, 2015ApJ...809...93D,  2016MNRAS.463L..22D, 2016ApJ...818...76J, 2018ApJ...866..110D}, the development of various kinds of instabilities in the disks ~\citep[e.g.,][]{2015ApJ...815L..15B, 2015ApJ...806L...7Z, 2015ApJ...813L..14B, 2016ApJ...821...82O,2009ApJ...697.1269J, 2014ApJ...796...31B, 2014ApJ...794...55T, 2018A&A...609A..50D}, a large
scale vertical magnetic field \citep{2018MNRAS.477.1239S} or with the destruction of large bodies in collisions \citep{2019ApJ...887L..15D,2020MNRAS.495..285N}. In all these papers, the source of disturbance is in the disk itself.

In our paper we discuss an alternative scenario for the formation of the observed structure. We investigate in the first time the dynamical response  of circumstellar disk on the perturbation associated  with the  clumpy accretion events. Using  the hydrodynamic simulations we calculate the disk images  at $1$ mm and discuss the results  in the context  with interferometric  observations  of  the protoplanetary disks with  ALMA.  

\section{Initial condition}
A model consists of a young star of solar mass ($ M_{\ast} =
M_{\odot} $) embedded in a gas disk with total mass is $ M_{disk}
= 0.01M_{\odot} $. At the beginning of simulation the disk matter was distributed
azimuthally symmetrical within the radii $ r_{in} = 0.5$ and
$r_{out} = 50$ AU. The initial density distribution of the disk
is
\begin{equation}
\rho(r,z,0)=\frac{\Sigma_0}{\sqrt{2\pi}H(r)}\frac{r_{in}}{r}e^{-\frac{z^2}{2H^2(r)}},
\end{equation}
where $\Sigma_0$ is arbitrary scale parameter, which is determined by
disk mass. Hydrostatic scale height is $H(r)=\sqrt{\frac{\kappa T_{mid}(r) r^3}{GM_{\ast} \mu m_H}}$, where $\kappa$, $G$ and $m_H$ are
the Boltzmann constant, the gravitational constant and the mass of
a hydrogen atom and $\mu=2.35$ is the mean molecular weight
\citep{1994A&A...286..149D}. Following~\citet{1997ApJ...490..368C}
we determine the law of midplane temperature distribution
$T_{mid}(r)=\sqrt[4]{\frac{\gamma}{4}}\sqrt{\frac{R_{\ast}}{r}}T_{\ast}$,
where $\gamma=0.05$ \citep{2004A&A...421.1075D}. The calculations were performed in the local thermodynamic equilibrium approximation  $P(r,z,t)=c^2(r)\rho(r,z,t)$, where $P$ is local pressure at the moment $t$ and $c$ is a sound speed. It was assumed that in the vertical direction along $z$ the disk is isothermal. The temperature of the star was assumed to be $T_{\ast} = 5780$ K and star radius is $R_{\ast}=R_{\odot}$. The disk relaxed during $600$ years, and then the remnant of the fallen gas clump was added into it.

\subsection{Impulse approximation to the clump-disk collision}

When a clump falls onto a disk, part of its kinetic energy is converted into thermal energy. An infrared spot should appear on the image of the disk where the clump fell. The thermal relaxation time in protoplanetary disks at the distance $\geq20$  AU from the star is much less than the local orbital period \citep{2017A&A...605A..30M}.  Therefore, the remnant of the fallen clump quickly comes to thermodynamic equilibrium with the matter of the disk.  It participates in the Keplerian motion of the matter, while maintaining the residual velocity component orthogonal to the plane of the disk. 

\begin{figure}[ht!]
\plotone{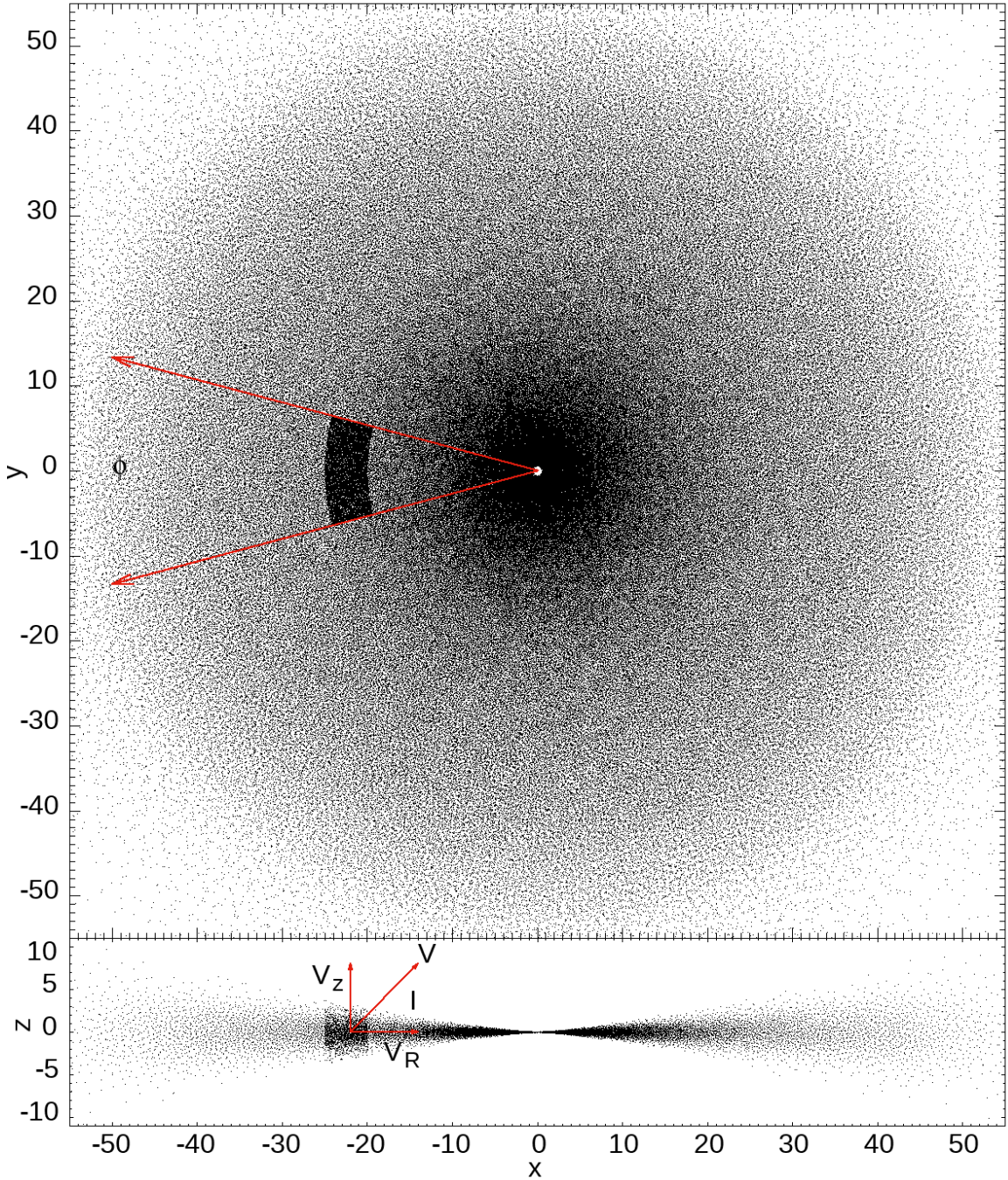}
\caption{\normalsize Particle distribution at the initial moment of the remnant of the clump motion. Top is the projection onto the plane of the disk, bottom is a section along the axis $y$. \label{fig:disk}}
\end{figure}

At the initial moment of time, we assume that the remnant of the clump has already reached thermodynamic equilibrium with the matter of the disk. The remnant was generated as a density perturbation on the disk in the form of a disk segment bounded by radii $R_0$ and step $dR$ and distributed over the azimuthal angle $\phi$ with the axis of symmetry along negative part of x-axis ($\phi=30^{\circ}$ for all models). The density of matter in the disturbance exceeded the local density of the disk by a factor of $\displaystyle K=\frac{\Sigma_{cl}}{\Sigma_d}$, where $\Sigma_{cl}$ and $\Sigma_d$ are local surface density of remnant and disk respectively. The remnant moves prograde. The particle velocity of the remnant is equal to  $\displaystyle V(R)=L\cdot V_k(R)$, where $V_k(R)$ is Keplerian velocity at a given distance from the star and $L$ is parameter of the problem. The velocity vector had a residual inclination to the disk plane $\displaystyle sin(I)=\frac{V_z(R)}{V(R)}$ (Fig.~\ref{fig:disk}). The residual angle of inclination of the remnant depends on the initial angle of the fall and the amount of kinetic energy that is spent on heating the disk in the region of fall. We considered a number of the possible options for the value and inclination of the velocity vector. The parameters of all calculated models are given in the Table~\ref{tab:models}.

\begin{figure*}[ht!]
\plotone{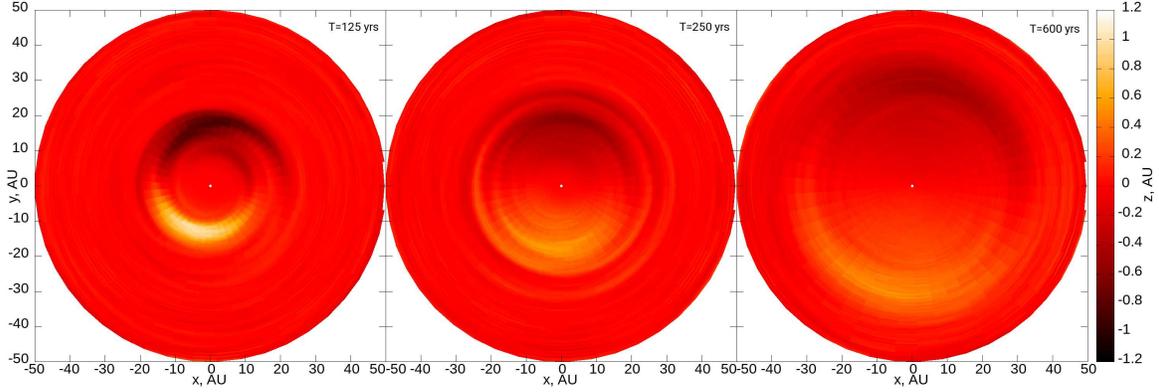}
\caption{\normalsize  
The average value of the $z$ coordinate of the particles in the cells of $R$, $\phi$. The model parameters are $K=3$, $I=30^\circ$ and $L=0.8$.  The time in years is in the upper right corner of the pictures. \label{fig:08inclImg}}
\end{figure*} 

\begin{deluxetable*}{ccccccccc}
\tablenum{1}
\tablecaption{The models parameters \label{tab:models}}
\tablewidth{0pt}
\tablehead{
\colhead{L} & \colhead{I} & \colhead{K} & \colhead{R} & \colhead{dR} & \colhead{$\phi$} & \colhead{Remnant mass} & \nocolhead{} & \colhead{Lifetime}\\
\colhead{Float} & \colhead{Degrees} & \colhead{Number} & \colhead{AU} &
\colhead{AU} & \colhead{Degrees} & \colhead{Jupiter mass} & \colhead{Sructures} & \colhead{yrs}
}
\decimalcolnumbers
\startdata
0.8 & 5 & 3 & 20 & 5 & 30 & 0.11 & Arc, One-hand spiral, Horseshoe & $> 600$ \\
0.8 & 10 & 3 & 20 & 5 & 30 & 0.11 & Arc, One-hand spiral, Horseshoe & $> 600$\\
0.8 & 20 & 3 & 20 & 5 & 30 & 0.11 & Arc, Faint two-arm spiral, Horseshoe & $> 600$\\
0.8 & 30 & 3 & 20 & 5 & 30 & 0.11 & Arc, Faint two-arm spiral, Horseshoe & $> 600$\\
0.8 & 10 & 5 & 20 & 5 & 30 & 0.19 & Arc, One-hand spiral, Horseshoe & $> 600$\\
\hline
1 & 5 & 3 & 20 & 5 & 30 & 0.11 & Arc, One-arm spiral, Multi Rings, Ring & $\sim 1000$ \\
1 & 10 & 3 & 20 & 5 & 30 & 0.11 & Arc, One-arm spiral, Ring & $> 600$\\
1 & 20 & 3 & 20 & 5 & 30 & 0.11 & Arc, One-arm spiral, Faint two-arm spiral  & $> 600$\\
1 & 30 & 3 & 20 & 5 & 30 & 0.11 & Arc, One-arm spiral, Faint two-arm spiral & $> 600$\\
1 & 5 & 5 & 20 & 5 & 30 & 0.19 & Arc, One-arm spiral, Multi Rings, Ring & $> 600$\\
1 & 5 & 10 & 20 & 5 & 30 & 0.38 & Arc, One-arm spiral, Multi Rings, Ring & $> 600$\\
\hline
1.2 & 5 & 3 & 20 & 5 & 30 & 0.11 & Arc, One-arm spiral, Multi Rings, Ring & $> 600$ \\
1.2 & 10 & 3 & 20 & 5 & 30 & 0.11 & Arc, Bright two-arm spiral & $> 600$\\
1.2 & 20 & 3 & 20 & 5 & 30 & 0.11 & Arc, Bright two-arm spiral & $> 600$\\
1.2 & 30 & 3 & 20 & 5 & 30 & 0.11 & Arc, Bright two-arm spiral, Asymmetric ring & $\sim 2000$\\
1.2 & 30 & 1 & 20 & 5 & 30 & 0.04 & Arc, Bright two-arm spiral & $> 600$\\
\hline
0.8 & 30 & 3 & 10 & 2 & 30 & 0.04 & Arc & $\sim 100$\\
1 & 30 & 3 & 10 & 2 & 30 & 0.04 & Arc & $\sim 100$\\
1.2 & 30 & 3 & 10 & 2 & 30 & 0.04 & Arc, Faint two-arm spiral, Multi Rings & $\sim 450$\\
\enddata
\tablecomments{The column of ``Structures'' lists the types of observed asymmetries in the order of their appearance in the disk images. The column of ``Lifetimes'' is a long-lived structures lifetime. }
\end{deluxetable*}

\begin{figure*}[ht!]
\plotone{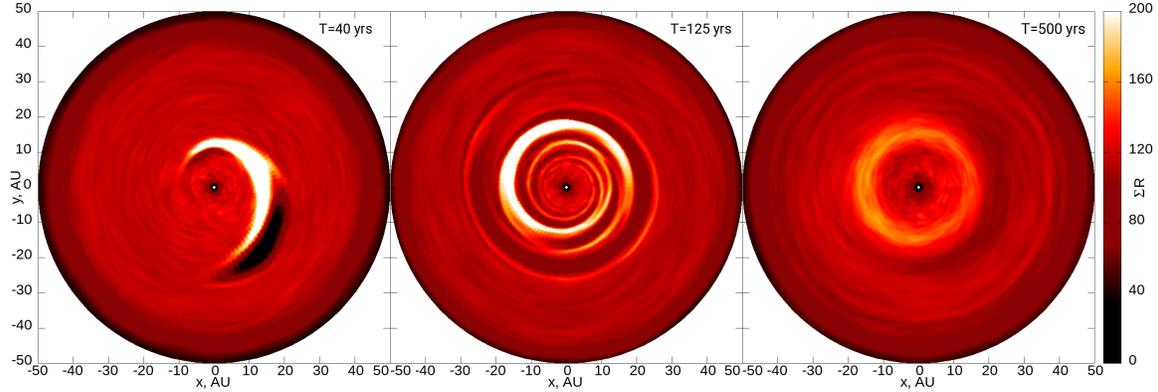}
\caption{\normalsize The surface density multiplied by the distance from the center of mass ($\Sigma R$). 
The model parameters are $K=3$, $I=10^\circ$ and $L=0.8$.  The time in years is in the upper right corner of the pictures. \label{fig:08sig}}
\end{figure*}

\begin{figure}[ht!]
\plotone{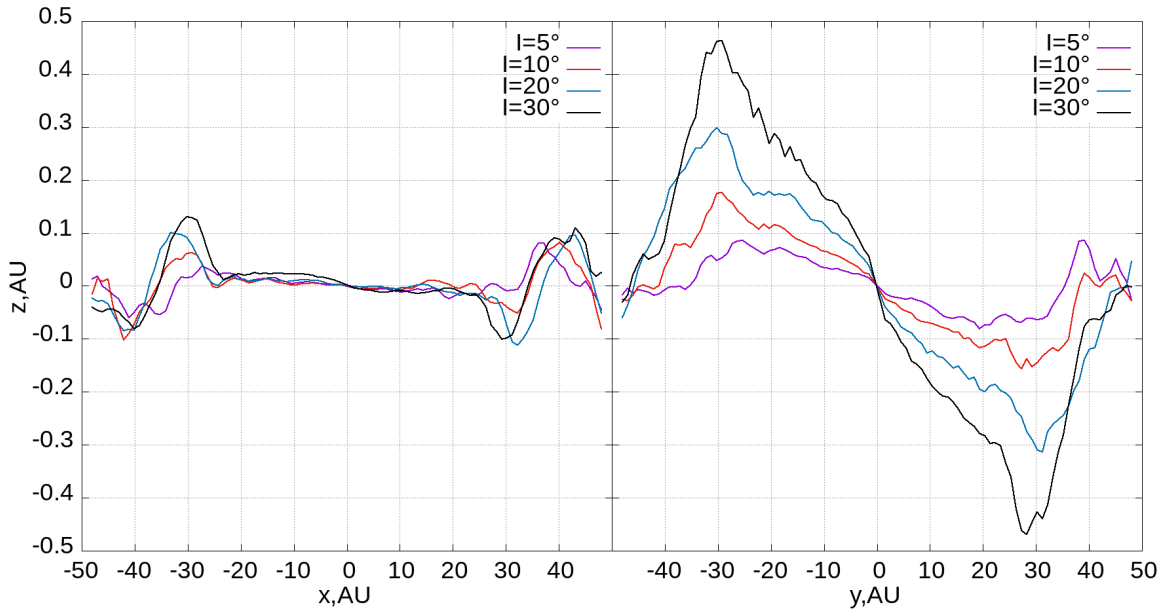}
\caption{\normalsize  
The average value of the $z$ coordinate of the particles in the cells of $R$, $\phi$ along $x$ (left) and $y$ (right) axes after $600$ years. The model parameters are $K=3$, $L=0.8$ and the angles are in the upper right corner of the pictures. \label{fig:08inclxy}}
\end{figure}
\begin{figure*}[h]
\plotone{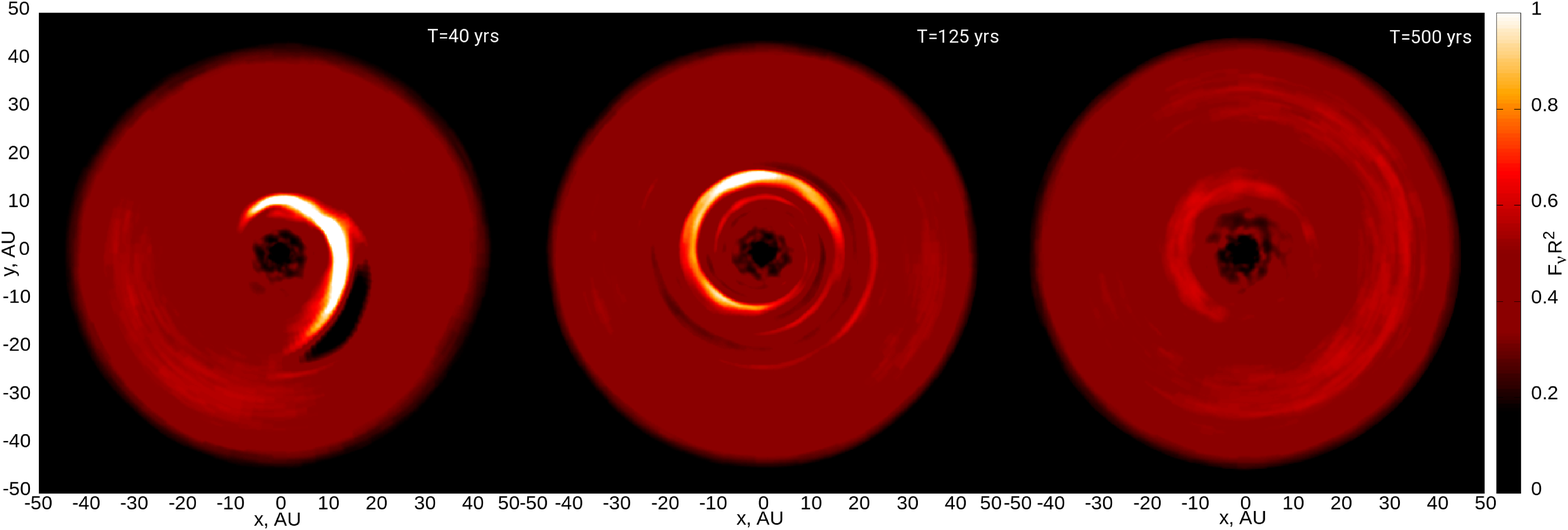}
\caption{\normalsize  
The images at wavelength 1 mm. The color shows the flux of radiation ($F_\nu$) multiplied by $R^2$ in conventional units. The model parameters are $K=3$, $I=10^\circ$ and $L=0.8$.  The time in years is in the upper right corner of the pictures. \label{fig:08img}}
\end{figure*}  

\section{Methods}
The evolution of the remnant in gas disk was simulated by
the SPH method (smooth particle hydrodynamics). The calculations
were performed using the code Gadget-2~\citep{2001NewA....6...79S,
2005MNRAS.364.1105S} modified by us \citep{2016Ap.....59..449D}. In total,
from $5\cdot 10^5$ to $2.5\cdot 10^6$ particles of the gas disk and from $5\cdot 10^3$ to $2\cdot 10^5$ particles of the perturbation were involved in the simulations. The calculations took into account the self-gravity of the disk. 

The simulated region was divided by $200\times30\times90$ cells in spherical coordinates ($R,\theta,\phi$), in which the average values of the SPH particle density were determined. We assume that dust particles with a size of 1, 10, 100 microns and 1 mm are well mixed with the gas, and are distributed according to the law$\frac{dn(s)}{ds}\propto s^{-3.5}$, where n is the concentration and s is the size of the dust grain \citep{1969JGR....74.2531D}. The dust to gas mass ratio in the disk was $0.01$ as the average in interstellar medium. The dust opacity is calculated using Mie theory for Magnesium-iron silicates~\citep{1995A&A...300..503D}. RADMC-3D~\citep{2012ascl.soft02015D} code was used for the 3-D radiative transfer calculations.  

\begin{figure*}[ht!]
\plotone{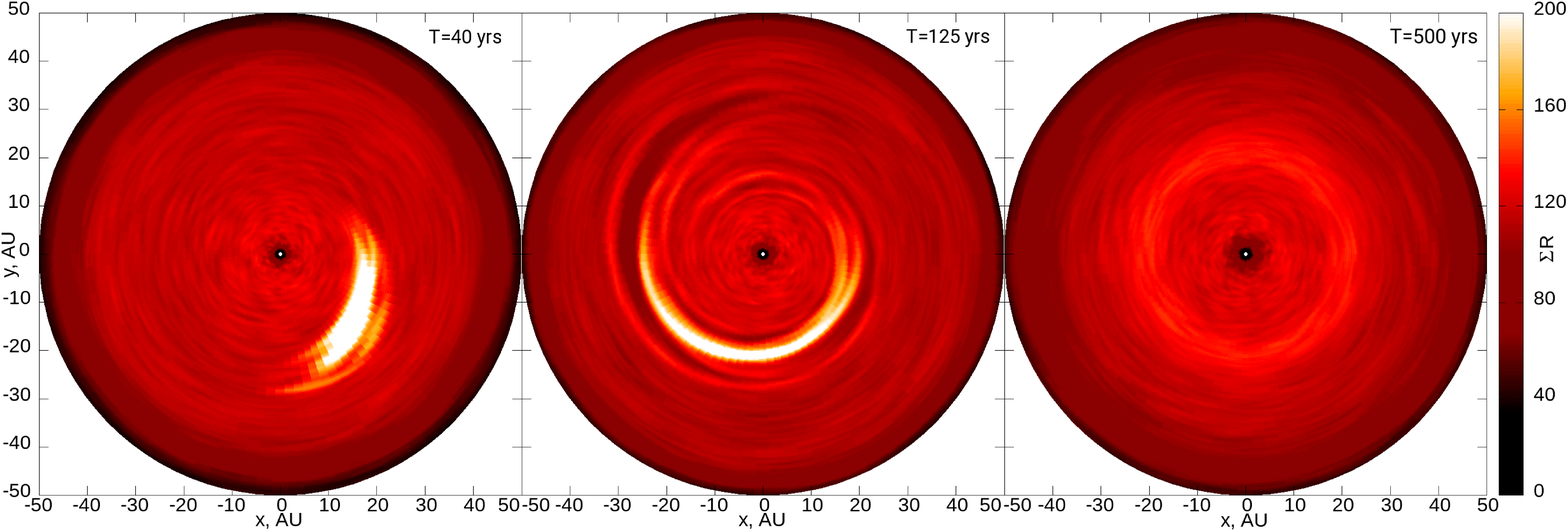}
\caption{\normalsize The same as in Fig.~\ref{fig:08sig} 
for model parameters $K=3$, $I=10^\circ$ and $L=1$. \label{fig:1sig}}
\end{figure*}
\begin{figure*}[ht!]
\plotone{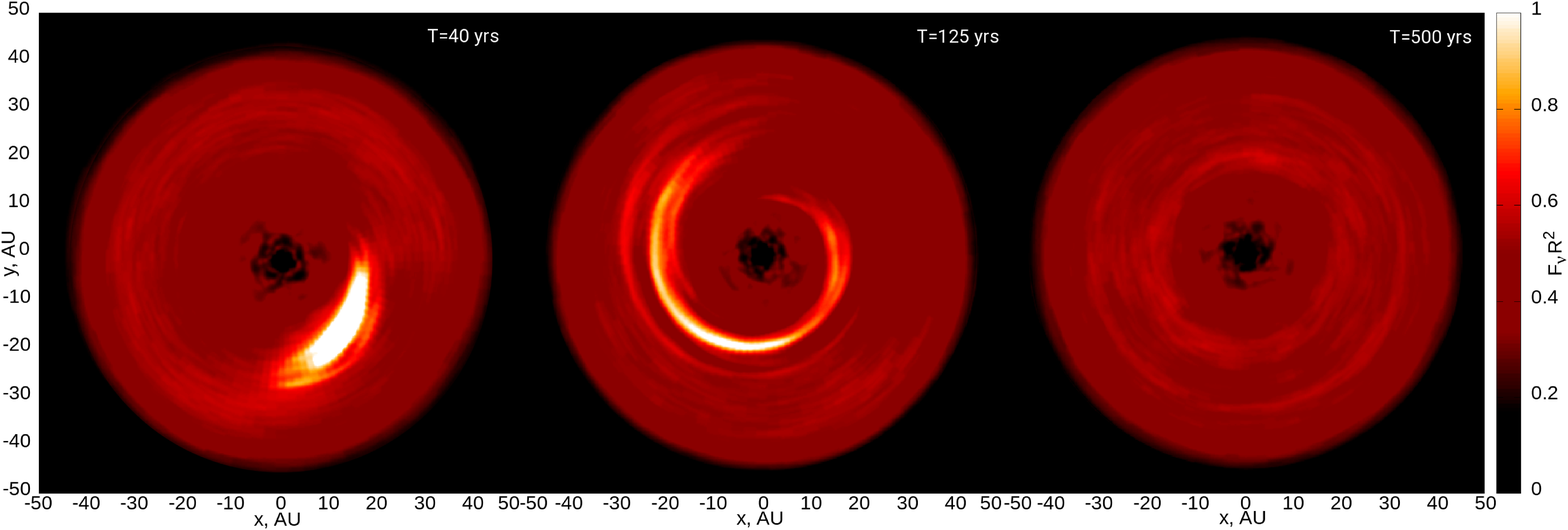}
\caption{\normalsize The same as in Fig.~\ref{fig:08img} 
for model parameters $K=3$, $I=10^\circ$ and $L=1$.  \label{fig:1img}}
\end{figure*}  
\begin{figure*}[ht!]
\plotone{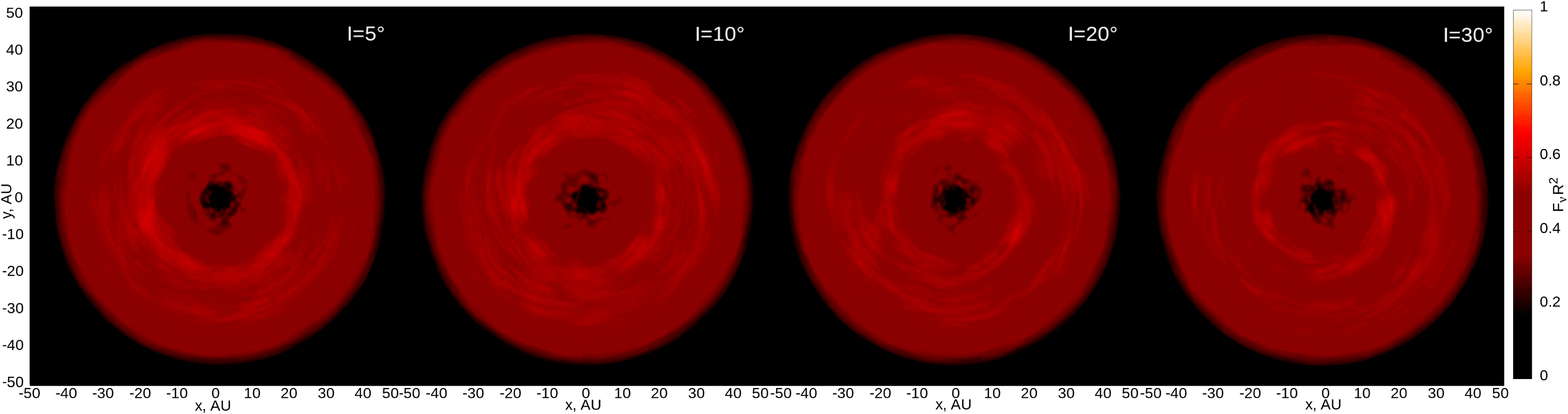}
\caption{\normalsize The same as in Fig.~\ref{fig:08img} 
for model parameters $K=3$, $L=1$ after $600$ years. The angle of $I$ is in the upper right corner of the pictures.  \label{fig:img600}}
\end{figure*} 
 \begin{figure*}[ht!]
\plotone{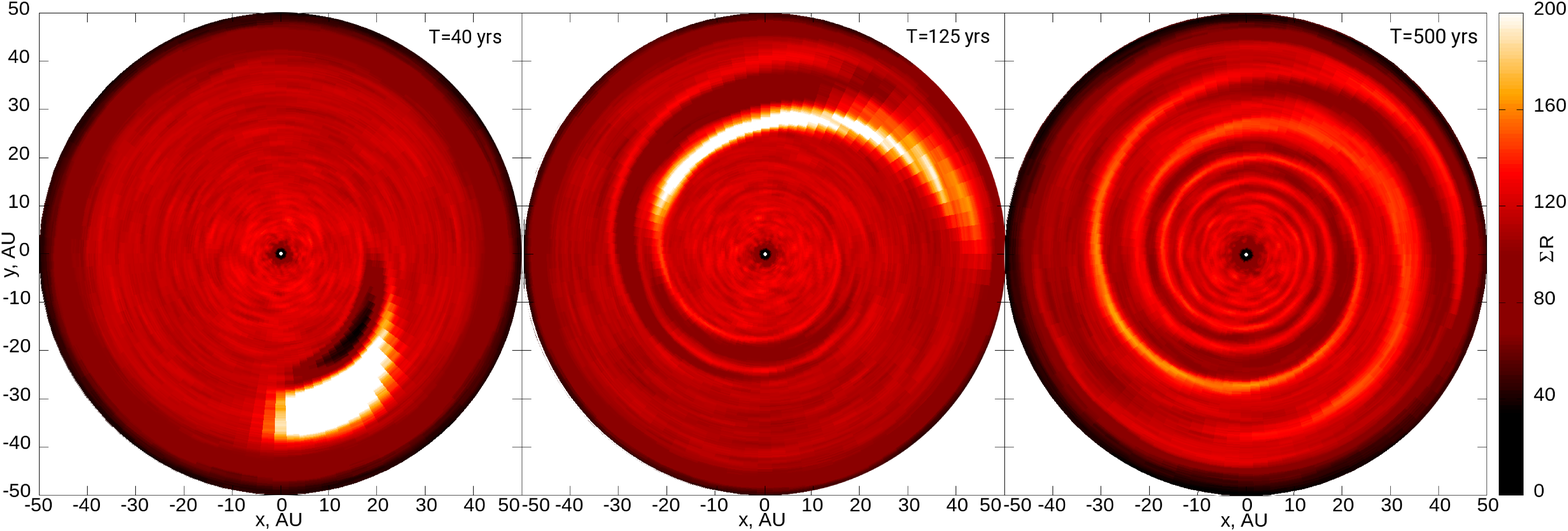}
\caption{\normalsize The same as in Fig.~\ref{fig:08sig} 
for model parameters $K=3$, $I=10^\circ$ and $L=1.2$. \label{fig:1.2sig}}
\end{figure*}
\begin{figure}[ht!]
\plotone{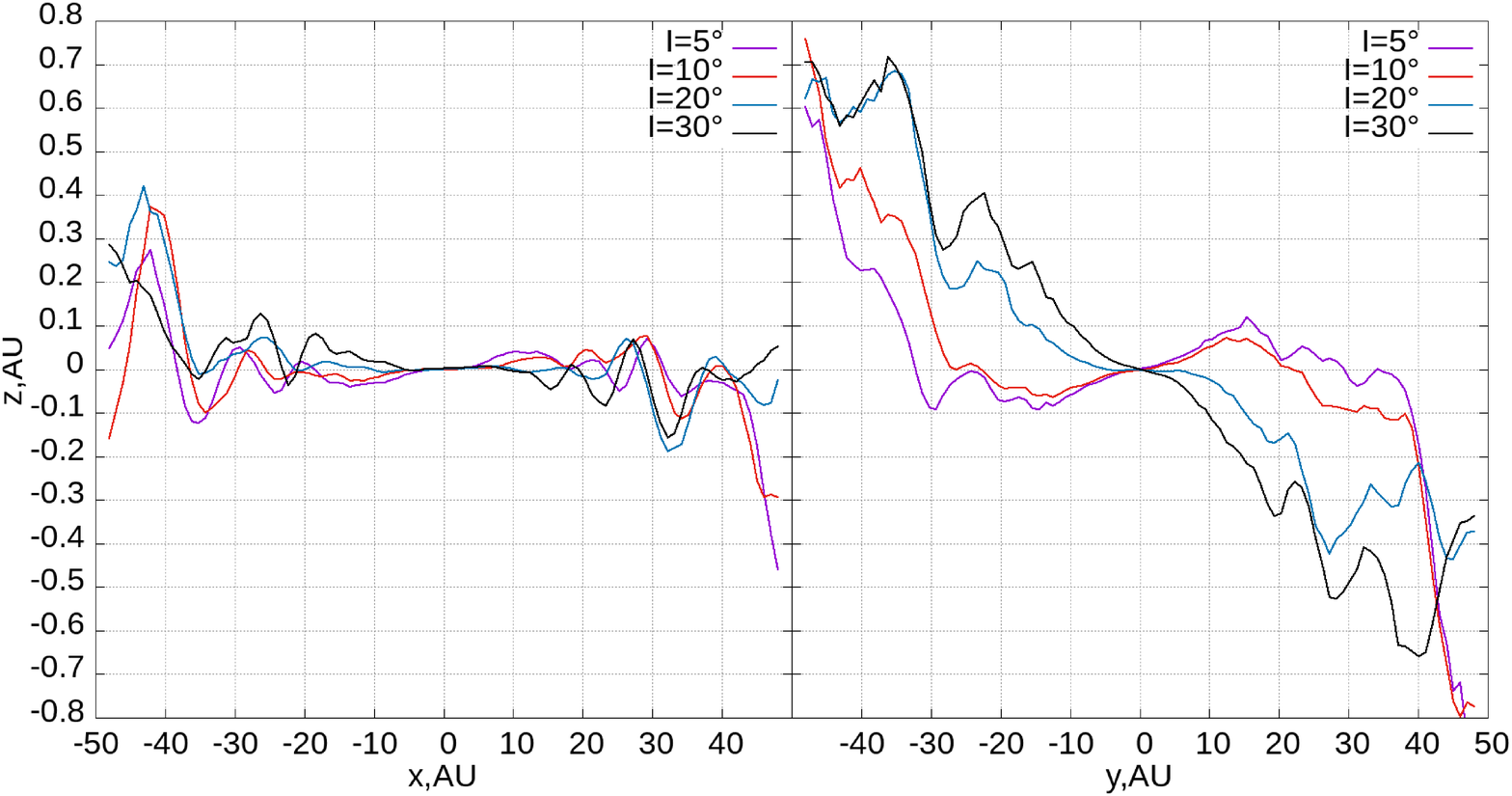}
\caption{\normalsize The same as in Fig.~\ref{fig:08inclxy} 
for model parameters $K=3$, $L=1.2$.\label{fig:12inclxy}}
\end{figure} 
\begin{figure*}[ht!]
\plotone{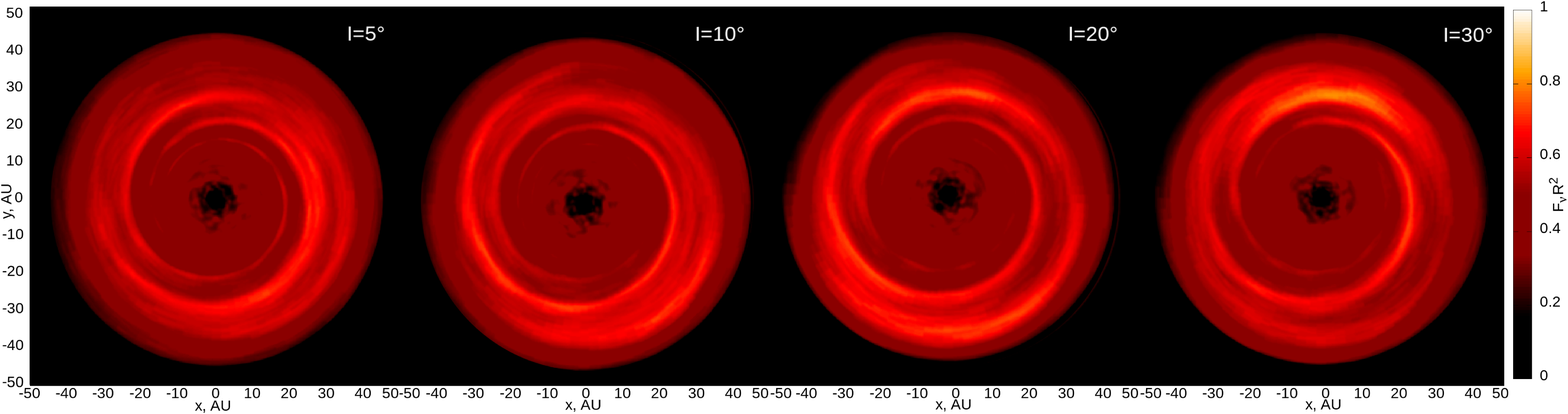}
\caption{\normalsize The same as in Fig.~\ref{fig:08img} 
for model parameters $K=3$, $L=1.2$ after $600$ years. The angle of $I$ is in the upper right corner of the pictures. \label{fig:img1_600}}
\end{figure*}  
\begin{figure}[ht!]
\plotone{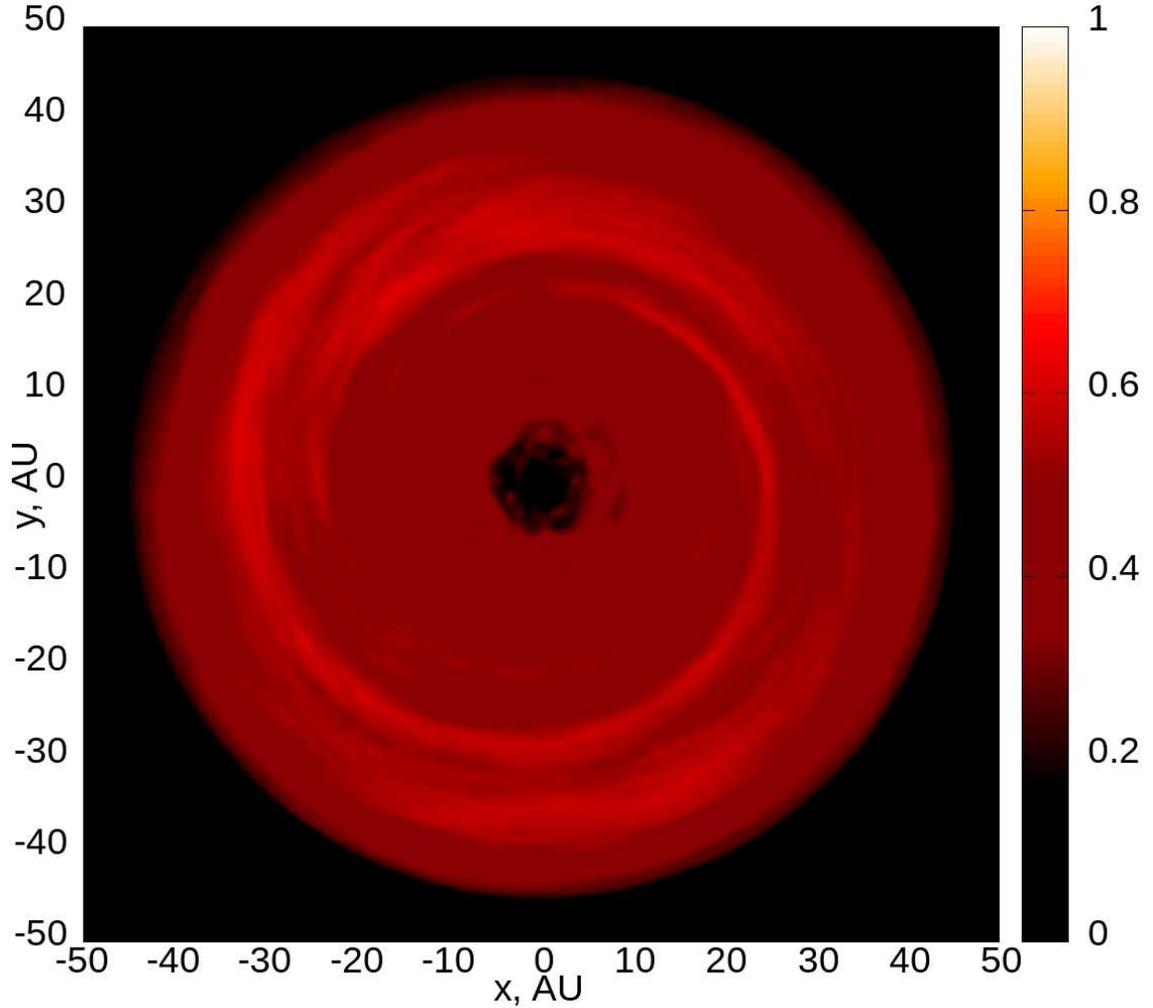}
\caption{\normalsize The same as in Fig.~\ref{fig:08img} 
for model parameters $K=1$, $I=30^\circ$, $L=1.2$ after $600$ years. \label{fig:K1I30}}
\end{figure}
\begin{figure*}[ht!]
\plotone{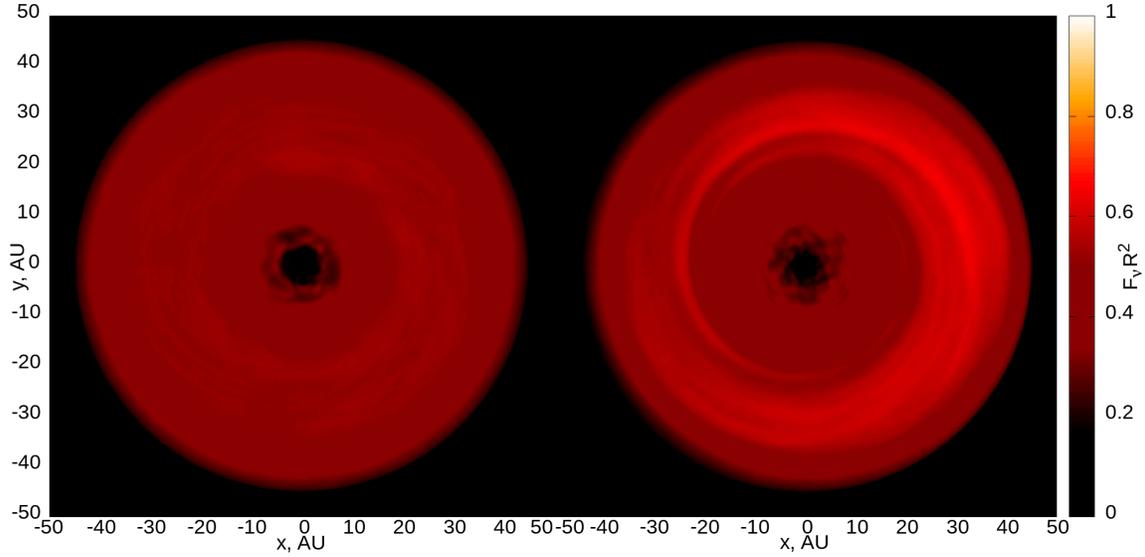}
\caption{\normalsize The same as in Fig.~\ref{fig:08img} 
for two models with $2.5\cdot 10^6$ particles. The models parameters $K=3$, $I=5^\circ$, $L=1$ (left) and $K=3$, $I=30^\circ$, $L=1.2$ (right) at the time $1000$ years. \label{fig:longImg}}
\end{figure*} 
\begin{figure}[ht!]
\plotone{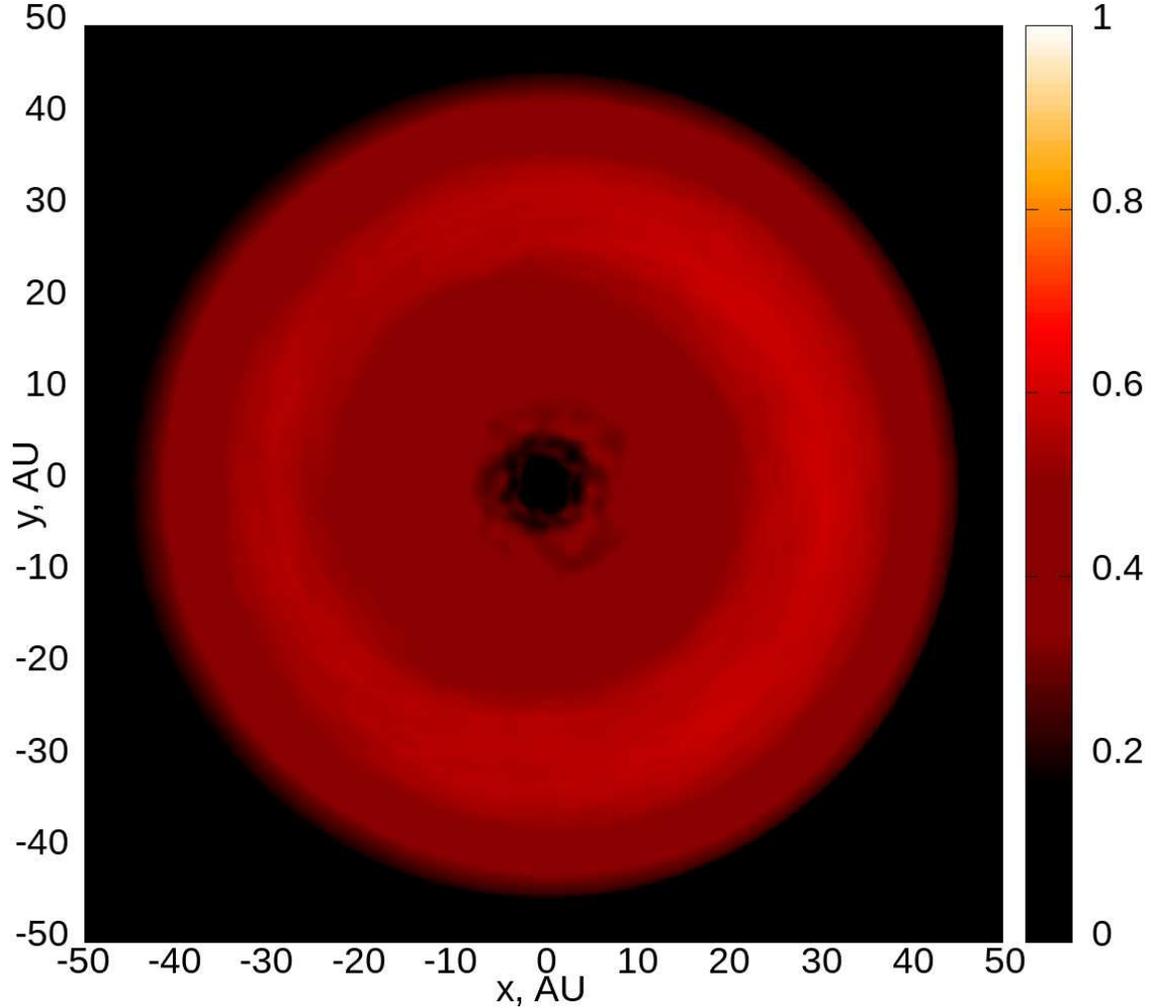}
\caption{\normalsize The same as in Fig.~\ref{fig:longImg} for model with parameters $K=3$, $I=30^\circ$, $L=1.2$ at the time $2000$ years. \label{fig:im2000}}
\end{figure} 
\begin{figure}[ht!]
\plotone{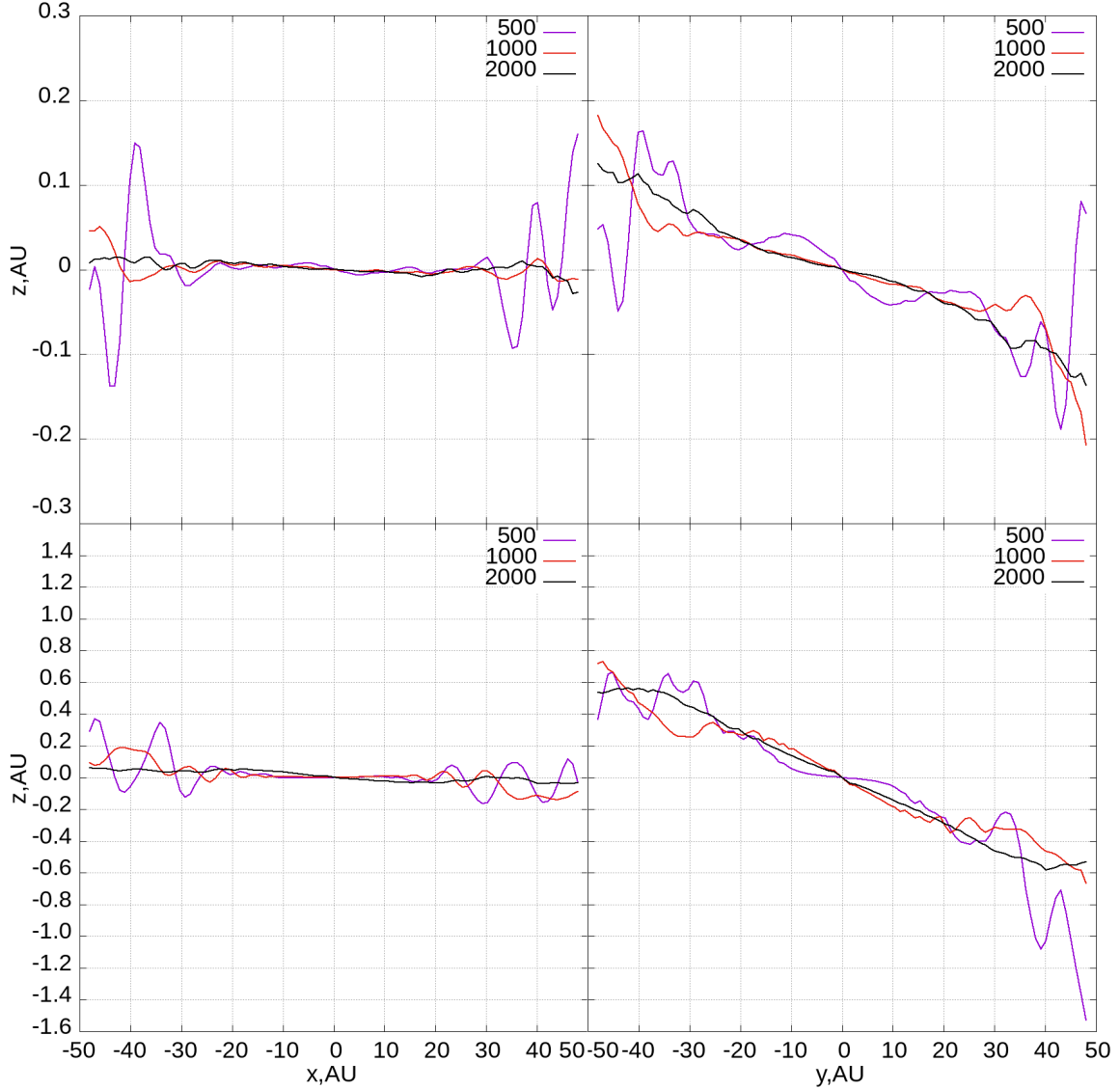}
\caption{\normalsize The same as in Fig.~\ref{fig:08inclxy} 
for models with parameters $K=3$, $I=5^\circ$, $L=1$ (top) and $K=3$, $I=30^\circ$, $L=1.2$ (bottom). The time in years is in the upper right corner of the pictures. \label{fig:longIncl}}
\end{figure} 
\begin{figure*}[ht!]
\plotone{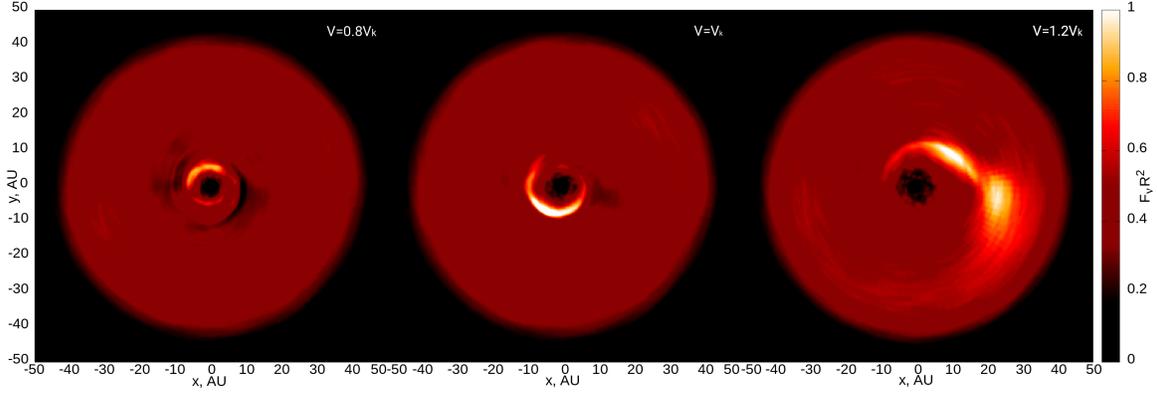}
\caption{\normalsize The same as in Fig.~\ref{fig:08img} 
for models with parameters $K=3$, $I=30^\circ$, $R_0=10$ AU and $dR=2$ AU after $36.5$ years. The parameter of $L$ is in the upper right corner of the pictures. \label{fig:closeImg}}
\end{figure*} 
\begin{figure}[ht!]
\plotone{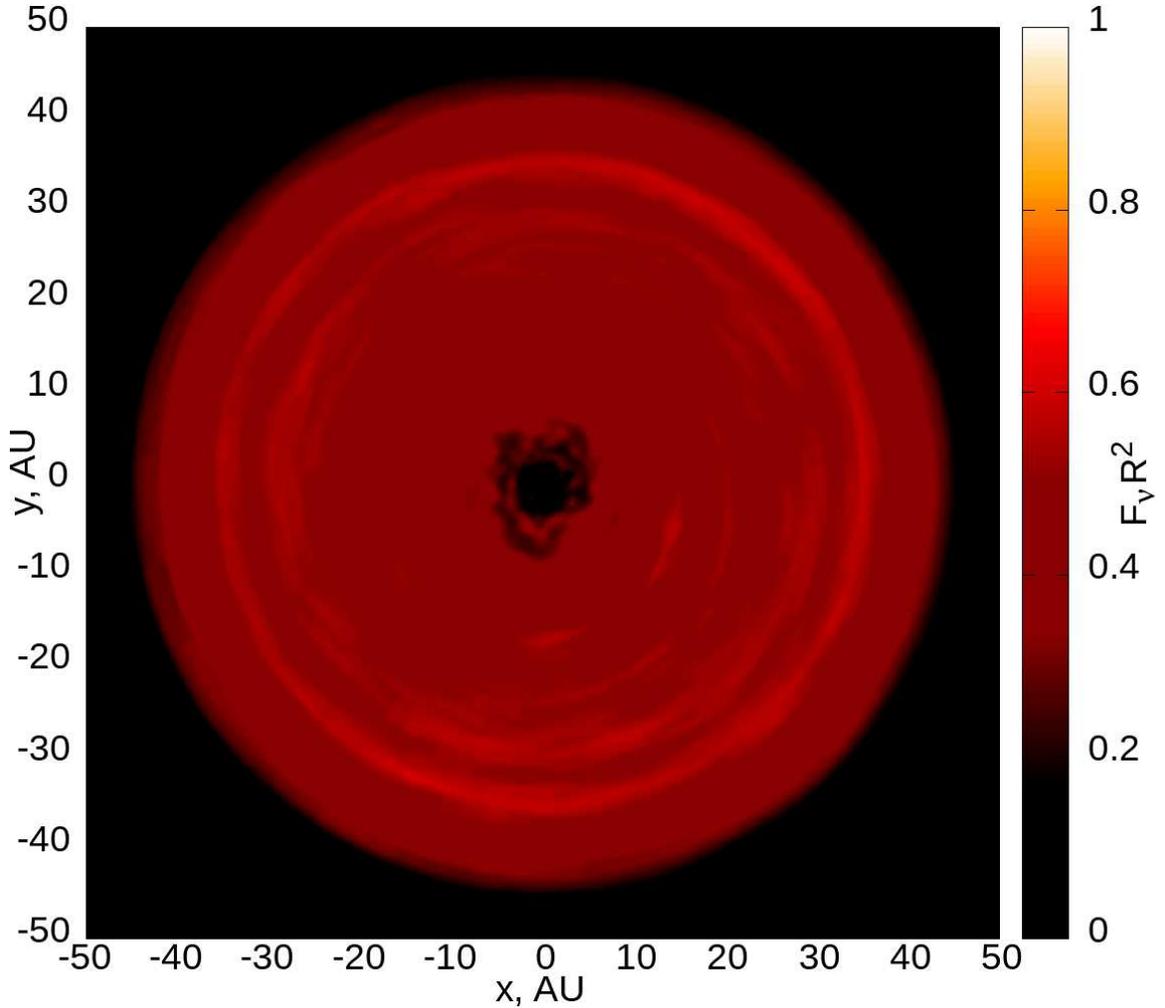}
\caption{\normalsize The same as in Fig.~\ref{fig:closeImg} 
for models with parameters $K=3$, $I=30^\circ$, $L=1.2$ for the time $400$ years. \label{fig:close400}}
\end{figure} 
\begin{figure}[ht!]
\plotone{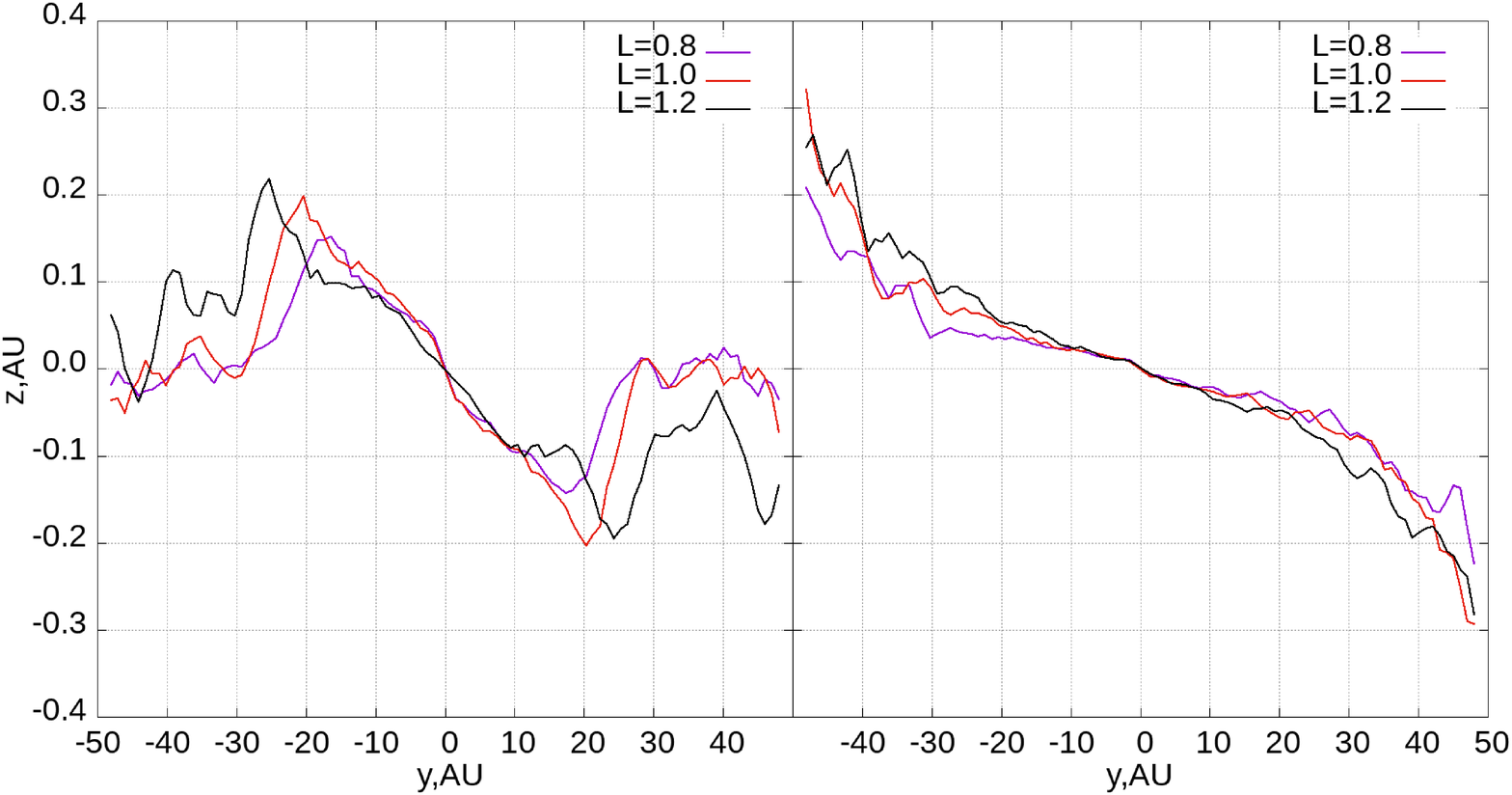}
\caption{\normalsize The average value of the $z$ along $y$ axes after $365$ (left) and $1460$ (right) years. The model parameters are $K=3$, $I=30$. The values of $L$ are in the upper right corner of the pictures. \label{fig:yincl}}
\end{figure} 

\section{Results} 
The emergence of the remnant of the clump of matter in the disk leads to the propagation of density waves in horizontal and vertical directions.  The strongest perturbations arise at large values of the parameters $K$, $L$, and $I$, as expected. Due to the differential rotation of the Keplerian disk the remnant stretches and transform during the time.  A local increase in the surface density leads to the appearance of corresponding large-scale inhomogeneities in the disk images.

\subsection{Perturbations at large radii} 
For the models discussed here, the initial position of the disturbing remnant of the clump was set equal to $R_0 = 20$ AU with a step of $dR = 5$ AU. Calculations have shown that the parameter $L$, which characterizes the kinetic energy of the falling clump, has the greatest influence on the type of disturbance in the disk. Therefore, we will sequentially discuss the three energy regimes considered in our models. 

\subsubsection{Sub Keplerian perturbations}
In this case, when the parameter is $L = 0.8$, due to the differential rotation of the Keplerian disk the remnant stretches and transform to a piece of arc reembling a cyclonic vortex and then turn into a spiral during one revolution around the star ($\sim 125$ years).  Since the matter of the disk is involved in its motion the spiral dips and thickening of the density are visible in the disk (Fig.~\ref{fig:08sig}). In the central part of the disk, the spiral splits into two branches. The spiral structure quickly ($\sim 300$ years) twists into an asymmetric ring, which, scattering in the disk, retains the asymmetry until the end of the calculations ($600$ years). This asymmetry rotates with the disk. 

The wave also passes in the vertical direction along the disk, which extends inside and to the edge of the disk. Perturbation twists the central plane of the disk. The maximum distortion of the disk plane occurs near the axis $y$, but does not coincide with its position. The inner parts of the disk inclines relative to the periphery. Over time, the radius of the outer boundary of the inclined area increases. It distorts 30 AU at the time of $600$ years  (Fig.~\ref{fig:08inclImg}). In this case, the tilt of the disk plane relative to the initial position is approximately equal to $0.2^\circ$ at $I = 5^\circ $ and $0.9^\circ$ in the case of $I = 30^\circ$. An increase in the angle of $I$ does not affect the speed of propagation of global perturbation to the edge of the disk (Fig.~\ref{fig:08inclxy}). Increasing the initial density (parameter K) of the clump increases the vertical distortion of the disk.  

The perturbations described above show themselves on the images of protoplanetary disks. The form of asymmetric structures in images corresponds to the perturbations of the density of the disk matter. However, on the periphery of the disk is visible a shadow from the matter above the disk plane. The asymmetric ring-shaped structure on the images has a horseshoe-shaped form (Fig.~\ref{fig:08img}). Calculations have shown that the minimum value of $K$ in which this structure is visible on the images, equal to $3$ (about $0.1$ of Jupiter mass). An increase in the parameter $K$ to $5 $ increases the brightness of the structure and their lifetime, but its horseshoe form is preserved.  

\subsubsection{Keplerian perturbations}
For this class of models of the phase of the disintegration of the clump remnant during the first orbit similar to the previous one (Fig.~\ref{fig:1sig}). A piece of arc is converted to the one-hand spiral. It twisted into a symmetric ring-shaped structure during the next few revolutions ($\sim 500$ years). 

An increase of the radius of distortion of the central plane of the disk relative to the initial position is faster than in the previous case. It reaches $40$ AU at the time $600$ years. The maximum inclination angle is $0.2^\circ$ in the case $I=5^\circ$ and $0.8^\circ$ at $I = 30^\circ$. The direction of maximum distortion in the vertical direction is also close to the axis $y$, but does not coincide with it.

The visible asymmetry on the images of the disk are also appropriate for surface density (Fig.~\ref{fig:1img}).  In this case, the propagation of the density wave gives multi-lane image of the protoplanetary disk at a certain point in time (right image of Fig.~\ref{fig:1img}). 
An increase in the angle of $I$ affects the image of the ring-shaped structure. Two symmetric weakly pronounced spirals are visible on the images instead of the ring, if the value of $I\geq20$ (Fig.~\ref{fig:img600}). 

\subsubsection{Super Keplerian perturbations}
In this case, the clump matter motion in the protoplanetary disk causes severe density perturbations and significantly distorts the disk in the vertical direction. As in previous cases, during one convolution of the clump, the vortex-like structure is stretched into a spiral, which is converted into two spirals during the next revolution (Fig.~\ref{fig:1.2sig}). Each of the spirals is logarithmic, they are shifted by phase relative to each other by $180^\circ$. The form of the spirals weakly depends on the angle of inclination of $I$. 

Disk distortion in the vertical direction differs from the models described above for this case. The periphery of the disk is distorted, and the inner parts of the disk deforms weaker. The waves are still propagating along the disk in a vertical direction at the time of 600 years as seen from Fig.~\ref{fig:12inclxy}. In this case, the distortion of the inner region depends on the inclination angle and has the opposite character for $I<20$ and $I\geq 20$.  

It become seen noticeable asymmetry of the spirals on the periphery of the disk in images with an increase in the angle of inclination of $I$ (Fig.~\ref{fig:img1_600}). 

For this class of models, the case $K = 1$ (corresponds to the mass of the remnant $\sim 12 M_{\oplus}$) was considered. In this case, the perturbation is weaker; however, the two-arm spiral can also be identified in the disk images. It is more pronounced at a larger inclination angle $I$ (Fig.~\ref{fig:K1I30}). 

\subsubsection{Long-term dynamics} 
For the models described above, the number of SPH particles was $5\cdot 10^5$. These models have a lower resolution compared to the models, the calculations of which involved $2.5\cdot 10^6$ particles. However, the calculations showed that at the initial phases of clump destruction, the images of disks obtained on the basis of models with a small number of particles show the same structures as for more accurate models. However, to study the long-term dynamics of the fallen clump remnant, higher resolution is required. We have calculated two limiting cases that correspond to the parameters that cause the minimum ($K=3,I=5,L=1$) and maximum ($K=3,I=30,L=1.2$) disturbance in the disk.  

Over time, density waves scattered and the all structure settles down to the plane of the disk. The ring of the first model stretches along the radius and loses brightness mixing with the matter of the protoplanetary disk with time. It is faintly noticeable at $1000$ years after the fall of the clump (Fig.~\ref{fig:longImg} left). Than it is not visible against the background of  the disk matter. For the second model spiral waves are still noticeable after $1000$ years (Fig.~\ref{fig:longImg} right), but after $\sim 2000$ years they completely disappear. An asymmetric ring structure can be seen on the disk by the time of 2000 years (Fig.~\ref{fig:im2000}). 

The charateristic time of the disk dynamic relaxation after the fall of the clump also depends on the place of its fall. For example, with $R_0 = 30$ AU and the same clump parameters as in the previous model, the lifetime of spirals and ring structures generated by its fall increases to $4 \times 10^3$ years. At $R_0 = 50$ AU, the characteristic relaxation time of disturbances on the disk is even longer: $\sim 10^4$ years.

On the Fig.~\ref{fig:longIncl} it is seen that the plane of the disk settles to its original position over time. However, even 2000 years after the fall of the clump, a slight inclination of the disk plane near the axis $y$ remains. For the first model, it is equal to $\sim 0.14^\circ$, and for the second, it is $\sim 0.72^\circ$. 

\subsection{Perturbations at small radii} 

In this class of models the perturbation was located near the star at the distance $R_0 = 10$ AU with the step of $dR = 2$ AU. The clump had parameters $\phi=30^\circ$, $K=3$ and $I=30^\circ$. The parameter $L$ was varied. The mass of the clump was about $13 M_{\oplus}$. 

Fig.~\ref{fig:closeImg} shows images of the disk for three models with the L parameter $0.8$, $1$ and $1.2$ after one period ($36.5$ years). Bright dense structures and areas of shadow, which are caused by matter rising above the plane of the disk, are visible on the disk. But all structures are scattered after next period for models with $L \leq 1$. For case $L=1.2$ waves propagate along the disk, which in time from 250 to 500 years can be seen in the images as a multi-lane structure (Fig.~\ref{fig:close400}). 

In the vertical direction, the disk is distorted in all considered cases. The maximum distortion is achieved near the $y$ axis as for models distant from the star. The Fig.~\ref{fig:yincl} shows the average values of $z$ along the y-axis for two points in time $365$ and $1460$ years. One can see in all cases, the perturbation propagates outward from the inner part of the disk, tilting its central plane. With an increase in L, the final inclination of the disk increases, but remains within $0.5^\circ$. 

\section{Conclusion} 
Calculations have shown that at the initial stages of the remnant of the clump disintegration, the structures that are visible in the images of the protoplanetary disk are similar for the entire set of the models. However, the shape of the final long-lived structure primarily depends on the kinetic energy of the falling clump. 

During the first revolution of the center of the remnant (at the initial moment of time), an arc-like structure resembling a vortex is visible in the disk image. Similar structures are observed, for example, in objects HD 135344B \citep{2018A&A...619A.161C} and HD 143006 \citep{2018ApJ...869L..50P}. At the next stage of evolution, the image shows a tightly wound spiral, as, for example, in the case of object HD 163296 \citep{2018ApJ...869L..42H}. In the case when the residual velocity of the remnant does not exceed the Keplerian velocity, a ring is a long-lived structure, which can also be asymmetric. In this case, at a certain moment in time, the passage of a wave over the disk can give a multi-lane structure in the image. The ring-shaped structure is visible in the images of a number of objects, for example, HD 169142 \citep{2017A&A...600A..72F}, HD 97048 \citep{2017A&A...597A..32V}, RU Lup, Elias 24, AS 209, GW Lup \citep{2018ApJ...869L..42H}. In the case of a high kinetic energy of a clump, a two-armed spiral appears on the disk image, which, after several thousand years, transforms into an asymmetric ring. Two-arm spirals were obtained on images of objects Elias 27, IM Lup, WaOph 6 \citep{2018ApJ...869L..43H}. The median age of these sources is $\sim 1$ Myr~\citep{2018ApJ...869L..42H}, and the youngest  sources have  estimated ages about a few tenths of Myr~\citep{2018ApJ...869L..43H}.

Since the velocity vector of the remnant has a residual inclination relative to the plane of the disk, it is distorted. However, over time, the inclination of the plane of the disk relative its original position decreases. The tilt angle at the end of calculations does not exceed $1^\circ$. Probably, for a more noticeable change in the inclination of the disk, a large mass of the clump remnant is required. In this work, we were looking for the minimum mass of the remnant, which develops into a structure visible in the image of the protoplanetary disk. It turned out that in the case of a high-energy fall, the minimum mass of the remnant of the clump is $\sim 10 M _{\oplus}$. If the residual velocity of the remnant does not exceed Kepler one, then the minimum mass is $\sim 0.1M_J$. 

Ring-shaped structures formed from the material of the remnant of the fallen clump and the protoplanetary disk are long-living structures and can exist for more than $3000$ years. Thus with a sufficiently large mass of the falling clump, the evolution of the disturbance can lead to the formation of a planet on an inclined orbit. It should be noted that Toomre parameter in the disks under consideration is equal to $\sim 40$  at a distance of $20$ AU. Consequently, in a more massive disk, the fall of a clump several times denser than the material of the disk can trigger the process of gravitational collapse and the formation of a planet.

The fall of the clump near the star can cause not only a FUOR flare, but also a strong increase in circumstellar extinction, leading to a deep and prolonged weakening of the optical brightness of the star and an increase in its infrared radiation. Such events were observed in three young objects: V1184 Tau~\citep{2009AstL...35..114G}, RW Aur~\citep{2015IBVS.6143....1S,2019A&A...625A..49K} and AA Tau~\citep{2021AJ....161...61C}. 

Let us make a rough estimate of the additional mass of circumstellar matter, which can be added to the mass of the disk during the lifetime of a star due to episodic falls of clumps. Suppose that the average age of stars with protoplanetary disks $10^6$ years~\citep{2018ApJ...869L..42H} and the average lifetime of one disturbance $3 \times 10^3$ years. This will give the probability of observing one event $P_1 \sim 3\times 10^{-3}$. The probability close to unity will be obtained when $\sim 3\times 10^3$ clumps fall during the disk lifetime ($\sim 1$~Myr). If the clump mass of $0.1 M_J$ is necessary to create a strong disturbance (see above), the disk mass will increase during this time by $0.03 M_\odot$ at an average accretion rate on the disk of $3\times 10^{-8} M_\odot yr^{-1}$. Such an increase in mass is not critical for a typical disk mass of $0.01-0.2 M_\odot$ and an accretion rate onto a young star of $\sim 10^{-7}-10^{-8} M_\odot yr^{-1}$.

Thus, the mechanism of the clumpy accretion in protoplanetary disks can explain the formation of the main types of structures identified in images of protoplanetary disks. In addition, one clump falling at an angle onto a protoplanetary disk can produce multi-lane bright ring-shaped structures. So far, we have demonstrated here the fundamental possibility of obtaining the observed structures on protoplanetary disks in the model of the clumpy accretion. Appreciate the complexity of this process, it needs a more detailed consideration, taking into account the thermal regime in the perturbed region of the disk. 

{\bf Acknowledgments.} It is a pleasure to thank the referee for valuable and useful remarks. Authors acknowledge the support of Ministry of Science and Higher Education of the Russian Federation under the grant 075-15-2020-780 (N13.1902.21.0039).

\software{Gadget-2~\citep{2001NewA....6...79S,
2005MNRAS.364.1105S}, RADMC-3D~\cite{2012ascl.soft02015D}}

\bibliography{DG21}{}
\bibliographystyle{aasjournal}



\end{document}